\documentclass[a4paper,11pt]{article}

\usepackage[colorlinks,urlcolor=blue,linkcolor=blue,citecolor=blue]{hyperref}
\usepackage{hyperref}
\expandafter\def\expandafter\UrlBreaks\expandafter{\UrlBreaks\do\/\do\*\do\-\do\~\do\'\do\"\do\-}
\usepackage{graphicx}
\usepackage{tabularx}
\usepackage{adjustbox,lipsum}
\usepackage[top=1.25in, bottom=1in, left=1in, right=1in]{geometry}
\usepackage{array}
\newcolumntype{P}[1]{>{\centering\arraybackslash}p{#1}}
\usepackage{multirow}
\usepackage{hyperref}
\usepackage{caption}
\usepackage{subcaption}

\newcounter{rowcntr}[table]
\renewcommand{\therowcntr}{[P\arabic{rowcntr}]}

\newcolumntype{N}{>{\refstepcounter{rowcntr}\therowcntr}c}

\AtBeginEnvironment{tabular}{\setcounter{rowcntr}{0}}

\begin{document}
\setcounter{secnumdepth}{3}
\title{Securing Bystander Privacy in Mixed Reality While Protecting the User Experience}

\author{Matthew~Corbett$^1$, Brendan~David-John$^1$, Jiacheng~Shang$^2$, \\Y.~Charlie~Hu$^3$,
        and~Bo~Ji$^1$ 
}

\date{%
    $^1$Virginia Tech, Blacksburg, VA, USA\\%
    $^2$Montclair State University, Montclair, NJ, USA\\%
    $^3$Purdue University, West Lafayette, IN, USA\\[2ex]%
    \today
}

\maketitle

\begin{abstract}
The modern Mixed Reality devices that make the Metaverse viable require vast information about the physical world and can also violate the privacy of unsuspecting or unwilling bystanders in their vicinity. In this article, we provide an introduction to the problem, existing solutions, and avenues for future research. 
\end{abstract}
\section{Introduction}
As the promise of the Metaverse grows, more and more powerful devices are required to satisfy user expectations of the kinds of experiences that make an immersive world so enticing. These experiences can immerse the user in a completely digital reality (Virtual Reality or VR), a physical reality altered with digital information (Augmented Reality or AR), or anything in between. We consider the breadth of devices and displays between, but not including, the unaltered physical world and the completely virtual world to be Mixed Reality (MR) as in the seminal works from Milgram and Kishino~\cite{milgram1994taxonomy}. MR devices include sensor suites that provide camera, depth, audio, and eye-tracking information that are essential to displaying immersive content and enabling naturalistic interaction. 
However, the outward-facing sensor suite on these devices does not discern between the data required for its functionality and data that can be used in a way that violates the privacy of \emph{bystanders}, i.e., those surrounding the user who have not or cannot give consent for their information to be collected.  

These violations, real or perceived, can manifest serious consequences for bystanders and device manufacturers alike. As shown in a brief case study in Section~\ref{sec:GoogleGlassStudy}, Google's Glass wearable was a new and innovative MR device that was ultimately hamstrung for many reasons, including the device's perceived lack of bystander privacy protection. The devices were met with large-scale criticism from both the public (wearers were derisively called ``Glassholes'') and governments alike. These concerns created marketability issues and also spurred legislation in multiple countries to regulate and restrain such devices. Ultimately, privacy concerns, as well as marketing and cost issues, became the demise of this product.  

We define the gap between the expectations of privacy that bystanders demand and the level of privacy that an MR device can provide to be the \emph{Bystander Privacy Problem} or \emph{BPP}. Through our research, we believe that this problem is comprised of two main components. The first is the technical vulnerabilities present in modern MR devices. These vulnerabilities stem from coarse-grained permissions for third-party applications and the ability of these applications to offload raw sensor data and violate the privacy of bystanders. This first component creates the conditions for the second, the perception that these devices can invade bystander or user privacy. This component does not have to be founded in technical reality (e.g., an actual privacy vulnerability) but has also been shown to be a hindrance to the success of MR devices. As shown in our case study in Section~\ref{sec:GoogleGlassStudy}, the BPP can create real issues for bystanders and device manufacturers. Each component can exist without the other, but we believe that the existence of technical vulnerabilities can exacerbate the perception issues that MR devices have suffered from previously.

A simple and naive solution to this problem would be to strip all bystander data from any recording (audio, visual, etc.), however, certain applications could require such information to function. Consider an example of a facial recognition program, implemented on an AR device, to assist patients in a memory care facility. This application could seek to assist the patient (i.e., the user) with remembering the names of friends, family, and healthcare workers by detecting these faces and labeling them in 3D on the augmented display of the device. In this case, detecting the identity of bystanders (e.g., other patients) could be a privacy violation. Knowing that \emph{some} faces are required, we are then presented with the problem of how to discern which faces are to be presented, known as \emph{subjects},
and which are simply bystanders. This means that we cannot simply remove all identifying information, such as faces, from sensor input. Doing so would limit the functionality of legitimate third-party applications that may require the use of this information. We must decide which information to remove and which to provide to such applications.

In this work, we seek to expose new or experienced MR application developers and manufacturers to the BPP, providing context and considerations to shape the development of MR applications and future MR privacy solutions. Additionally, we further explore existing solutions to address the BPP, and where the solutions often fall short by removing required data for legitimate applications or inadequately addressing bystander privacy concerns.
Finally, we present potential future directions in this field, including research into technical solutions and ways to address potentially unfounded perceptions of privacy violations from bystanders. We intend to illuminate the issue of bystander privacy in MR devices, and specifically, the lack of viable solutions~(Section~\ref{Problems}), while presenting a framework for future solutions designed to address this problem (Section~\ref{AddressingIssues}). Finally, we have presented a chronologically-ordered table (Table~\ref{tab:Q2}) that roughly outlines the published work on the topic. While this list is not exhaustive, we believe that it is representative of the state-of-the-art in bystander protection in MR.

We have created this work to facilitate a better understanding of the BPP in the hope that technical and perceptual issues do not further hinder the adoption of these powerful and useful MR devices.

\section{WHAT ARE BYSTANDERS, USERS, AND SUBJECTS?}
\label{se:WhatisABystander}
As part of the introduction to the BPP, we present a few key definitions for clarity. A \emph{user} is a person who wears an MR device. A \emph{subject} is a person with whom the user intends to interact and has also given any form of consent for data capture. This definition can also include subjects in previous interactions who have given consent for the personal data to be used and can be used in subsequent interactions. If such a subject were to rescind their consent for this collection, they would revert to the protections offered to bystanders. Additionally works such as~\ref{ref:adjacentactorprivacy} make an even more nuanced approach to subjects, including so-called ``co-users'' or those who have some control over the application and gain benefit from it. An example of a ``co-user'' might be made during a structural bridge inspection, if one inspector is using AR to detect and localize a crack in the concrete bridge supports and a second inspector (the ``co-user'') is observing on a laptop-based video feed. Our use of the word ``subject'' includes all of these classifications into one over-arching term, simplifying the discussion. However, a more nuanced approach could include each of these classifications as a separate level of protection/access (discussed further in Section~\ref{FutureDirections}). 

\emph{Bystanders} are classified as any non-user, non-subject third-party surrounding the device during its use. This bystander can be aware or unaware of the device's presence. In the bridge inspection example above, a ``bystander'' would be a person walking by the inspection who is unintentionally captured by the MR device's sensors. We acknowledge that these definitions are relatively simple and that more nuanced and detailed definitions exist such as the taxonomy of Pierce et al. in~\ref{ref:adjacentactorprivacy}. We choose to use this simple and binary definition as it streamlines the decision between the two labels (i.e., subject and bystander), making generalization across many different contexts possible. Using a more subtle approach, with multiple definitions, could be overly burdensome when considering a system spanning across scenarios like interpersonal communications, industry, exercise and fitness uses, and others. 

\begin{figure}[tp]
     \centering
     \includegraphics[width=0.3\textwidth]{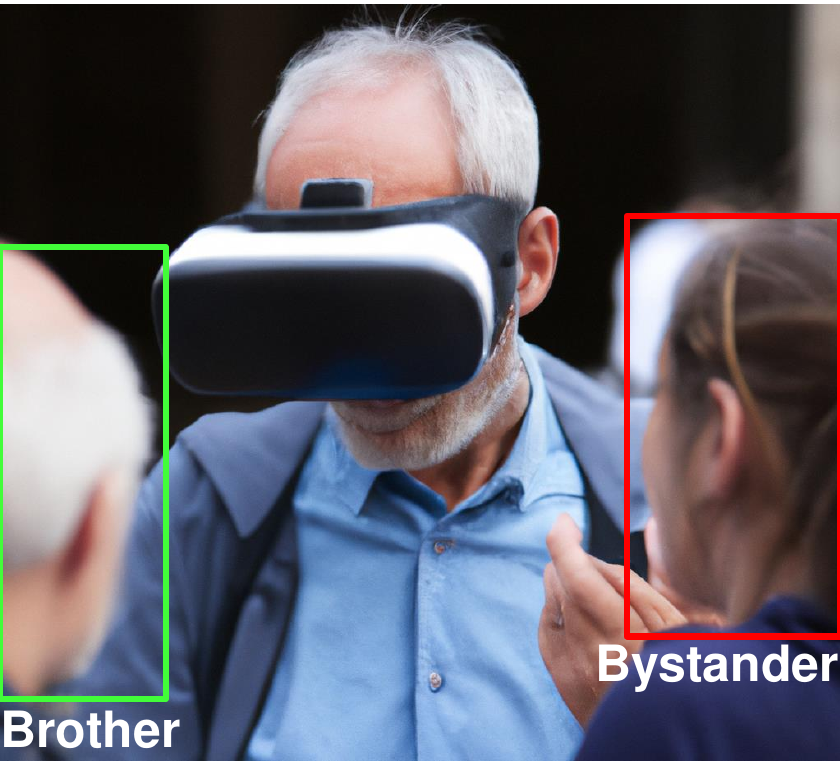}
     \caption{An illustration of a memory care use case for AR devices. In this example, a memory care patient requires assistance with identifying close friends and relatives. An application, using an AR device, identifies faces using facial recognition software and presents the names and/or titles of these persons in order to assist the user. In this example, the user is wearing the device and the subject is the person labeled as ``Brother''. The bystander, uninvolved with this interaction, is labeled by the red bounding box.
     }
             \vspace{-0pt}
    \label{fig:MedicalExample}
\end{figure}

In order to make these definitions more concrete, we present an example illustrated in Fig.~\ref{fig:MedicalExample}. In this example, a patient in memory care is suffering from memory loss from Alzheimer's disease. The patient requires assistance to identify close friends and family. The application, on an AR device, uses facial recognition to identify and label persons of interest to the patient in real-time. However, identifying bystanders, such as other patients or family members of other patients, would be a violation of privacy as their faces were used as part of the application without their consent. 

\section{MODERN MR DEVICES CREATE PRIVACY CONCERNS FOR BYSTANDERS}
Modern MR devices can effectively immerse the user in an altered reality. This immersion requires the use of sensors that can collect information about bystanders and violate privacy expectations. 
\label{Problems}

\subsection{The Bystander Privacy Problem}

\begin{table*}
\small
\begin{adjustbox}{max width=\textwidth}
\begin{tabularx}{\columnwidth}{|X||X|X|X|X|X|X|}
 \hline
 \textbf{Device} & \textbf{Released} & \textbf{Environ-ment Mapping} & \textbf{Cameras (including tracking cameras)} & \textbf{Eye Tracking Cameras} & \textbf{Depth Cameras} & \textbf{Microphone} \\
 \hline
  Microsoft HoloLens 2 & 2019 & Yes & 5 & Yes & Yes & Yes \\
 \hline
  Google Glass Enterprise Edition 2 & 2019 & No & 1 & No & No & Yes \\
 \hline
    Varjo XR-3 & 2020 & Yes & 2 & Yes & Yes & No \\
 \hline
 Magic Leap 2 & 2022 & Yes & 5 (including controller) & Yes & Yes & Yes \\
 \hline
Meta Quest Pro & 2022 & Yes & 5 & Yes & Yes & Yes \\
 \hline
  Xreal Air & 2022 & No & 0 & No & No & No \\
 \hline
 Meta Quest 3 & 2023 & Yes & 4 & No & Yes & Yes \\
 \hline
 Apple Vision Pro* & 2023 & Yes & 12 & Yes & Yes & Yes \\
 \hline
 
\end{tabularx}
\end{adjustbox}
\vspace{7pt}
\caption{ A list of popular MR devices released since 2019 and their onboard sensing capabilities.\\ (* = expected or assumed capabilities~\cite{AppleVisionPro})}
\label{tab:deviceListing}
\end{table*}

As explained in the Introduction, we believe the BPP is comprised of two parts, the technical vulnerabilities in MR devices and the perception that MR devices can violate bystander privacy. For the first part, we cite the Open Worldwide Application Security Project (OWASP) definition of a privacy violation. A technical vulnerability becomes a privacy violation when both private user information ``enters the program'' and this information is ``written to an external location''~\cite{OWASPPrivacy}. Research by Lehman et. al. in \ref{ref:hiddeninplainsight} has borne out this fear of technical vulnerabilities in MR devices. In so-called ``Hidden Operations'', a malicious application can collect camera frames, infer information such as gender, age, or the presence of specific objects, and exfiltrate that information using a network connection. If this occurs with sensitive information, we have satisfied OWASP's definition of a privacy violation. Violations like these are made possible through the relatively coarse-grained permissions granted to an application after a cursory prompt to the user.

With or without actual technical vulnerability, the perception problems caused by MR devices can be problematic. O'Hagan et. al. and Corbett et. al have shown that bystander perceptions of privacy in the face of these threats are generally negative~\ref{ref:understandingbystandersvaryingneeds}~\ref{ref:BystandAR}, especially before these bystanders are given information about how they may be protected by privacy solutions. These concerns have had tangible impacts on the devices, public perception, and international law. Public perception, specifically, has spurred legislation that is designed to control the risk these devices present to bystanders and has also contributed to the market failure of devices in the past. This consumer blowback and legislation threaten to derail the acceptance of MR technologies.



At the industrial level, even Facebook (now Meta) expressed concern about the use of facial recognition technologies on future devices, saying ``Face recognition $\dots$ might be the thorniest issue, where the benefits are so clear, and the risks are so clear, and we don’t know where to balance those things''~\cite{FacebookConsideringFacialRecog}. Quotes like this show both an acknowledgment and a lack of policy solutions to the bystander privacy problem.

\subsection{Modern MR Devices Are Uniquely Problematic}
\label{UniquelyProblematic}
An obvious comparison would be to evaluate the privacy concerns of MR devices as compared to smartphones. Both are capable of running MR applications and collecting information deemed sensitive to bystanders. However, some, such as the Information Technology and Innovation Foundation (ITIF), believe that MR devices are inherently more invasive than their smartphone counterparts~\ref{ref:PrivacyQuesationsExpansionAR}. When recording information (e.g., the front-facing camera is active), these devices give few clues to the purpose of this collection beyond non-salient visual indicators such as LED lights, or other passive warnings that have not been shown to adequately reduce bystander apprehension~\ref{ref:UserCetnricDesigns}. Recording visual data with smartphones normally involves pointing one of the embedded cameras at the target of the recording and revealing their behavior to the bystander within view. With wearable MR devices, the front-facing camera, depth sensor, and spatial audio microphones would be directed at the target when the user faces their direction. This changes the likelihood that the bystander would perceive that they are the subject of data collection and potentially being recorded. 

Finally, as shown in Table~\ref{tab:deviceListing}, modern MR devices contain far more sensors than their smartphone counterparts. Take for example the Samsung Galaxy 23 Ultra, the smartphone judged to be the largest and most capable camera array at the time of writing this paper. The Ultra has 4 photographic cameras, including rear and forward-facing cameras. The Apple Vision Pro is expected to have as many as 12 cameras onboard to accurately sense and understand the physical world~\cite{AppleVisionPro, BestSmartphoneCamera}. While the potential privacy threat of a device cannot be measured solely by the number of cameras, these sensors give MR devices a broad ability to collect data. 

This additional capability has been shown to make MR devices a more potent threat than their smartphone counterparts. Authors such as Hartzog et.al. and Dick have explained that this elevated threat is due to the ``always-on'' nature of MR devices, their mobility, and the difficulty a bystander may have in determining what is being collected~\cite{HatzogObscurity} \ref{ref:PrivacyQuesationsExpansionAR}. We believe that these capabilities set MR devices apart from smartphones, and create new and important research questions in order to protect bystander privacy.

\subsection{User-focused Privacy}
While this work is primarily concerned with protecting the privacy of bystanders in the field of view of an MR device, a more complete investigation of privacy issues in MR must include \emph{users} as well. Works such as \ref{ref:BANS} investigate how to protect the privacy of VR users in the presence of bystanders by alerting the user to the bystander's presence. These user-focused works provide a different approach to bystander privacy, in that they focus on protecting the user \emph{from} bystanders instead of protecting bystanders from the user (or more specifically, the user's device).

\subsection{Legislation and Global Policy}
In 1967, U.S. Supreme Court Justice Potter Stewart expressed that personal information cannot be considered private if the person ``knowingly exposes it to the public''~\ref{ref:KatzvUS}. Since this ruling, however, the definition of what can be considered ``reasonable'' has eroded as technology has advanced~\cite{DesertofUnreal}. As technology has enabled average users to infer more and more detailed (and potentially sensitive) information, what can be considered a ``reasonable'' expectation of privacy is under constant discussion and scrutiny. While these legal rulings and opinions are extremely useful in defining the context of the BPP, other work has sought to translate this into privacy frameworks and models. Such work posits that a person's right to privacy is based solely on the context at the moment of potential information capture, dictated by norms of ``flow'' and ``appropriateness''~\ref{ref:PrivacyAsContextualIntegrity}. Theories such as this would indicate that privacy solutions cannot be ``one size fits all'' and must rely on an understanding of the context in the moment of capture. Sensing a gap between theories about public privacy and protections and actual laws, exemplified by the advent of these new and more powerful devices and their ability to collect information, governments around the world have sought to fill gaps in policy with new and powerful legislation. 

As an example of such legislation, the General Data Protection Regulation (GDPR) in the European Union (EU), seeks to protect bystanders by requiring a
balance of data collection justification and bystander input to lawfully allow the collection of data. In the U.S., a ``patchwork of state and national policies'' comprise protections~\ref{ref:BalancingUserPrivacy}. Even
while different in content and scope, these laws emphasize consent in data collection from the person whose information is collected. This consent has been shown to make the collection of information far more
acceptable~\ref{ref:understandingbystandersvaryingneeds}, but such an option is rarely afforded to bystanders who are not even aware of the presence of the device. However, mechanisms to ``opt-in'' are far from standardized and generally only exist as academic solutions. Such mechanisms have not knowingly been implemented as part of any existing MR device and have only been proposed as potential future guidelines~\ref{ref:PrivacyQuesationsExpansionAR}.


\subsection{Manufacturer-Driven Policies and Standards}

Public concern over the potential of MR devices to violate bystander privacy has forced some device manufacturers to develop and enforce their own standards in an effort to ameliorate this concern. As an example, Apple has released details on privacy policies and mechanisms for its upcoming Vision Pro MR device~\cite{AppleVisionPro}. As mentioned previously in Section~\ref{UniquelyProblematic}, the Vision Pro has a large number of sensors, including 12 cameras, 6 microphones, and an array of eye-tracking sensors. This device, while not yet publicly available at the time of writing this paper, would be the most powerful MR device yet developed upon its release.

In order to lessen public concern, Apple has stated on the Vision Pro's website that ``Data from cameras and sensors is processed at the system level, so individual apps do not need to see your surroundings to enable spatial experiences.'' While we do not know exactly what this means, it appears to be a system similar to Darkly, that abstracts sensor data to give third-party applications an opaque view into the device's surroundings~\ref{ref:ScannerDarkly}. However, systems like Darkly provide an override for applications that legitimately need raw sensor data, like facial detection applications.
Apple has given no insight into how, or if, third-party applications would be able to access raw sensor data for purposes such as these. 


\subsection{Impacts on Future Devices}
As we will see with Google's Glass AR device, the perception that a device will negatively impact the privacy of bystanders in public places can hamper both the marketability of a device and run afoul of future regulations and legislation designed to protect such information. As public perception of the danger of collecting personal data in public places increases, device manufacturers must comply with ever-evolving regulations and codes, either officially codified into law or proposed as community best practices. Future solutions will need to address these concerns with consent mechanisms, information collection limits, and other methods to assure both a potential user base and potential bystanders that their devices are not actively collecting sensitive information. 

\section{GOOGLE GLASS: A CASE STUDY}
\label{sec:GoogleGlassStudy}
Google released the Glass in 2013 as one of the first attempts at a usable, wearable heads-up display.  While not truly an AR device, as the device did not interact and respond to the physical world, the Glass was designed to provide navigation and directions, notifications, and other passive alerts to the user. 



The Google Glass was met with an immediate negative perception, as indicated by widespread concerns about the cost, harassment in public,
and privacy. Specific to concerns about privacy, most concerns were due to its onboard camera~\cite{GoogleGlassCons}. The camera was capable of taking still images and video but required the user to either use a touchpad on the rim of the glasses or issue a voice command. Video capture was also signaled by a visible light on the glasses. This would prevent the completely stealthy capture of a photo or video by the user, and allow bystanders some knowledge of the user's intent. However, this did not stop local bars, advocacy groups, and local governments from protesting and banning the device from use. Users of the device were given the derisive term ``Glassholes'' for the perceived narcissism and intrusion into the privacy of others. Even the threat of facial recognition, a relatively new field ten years ago, rose to the surface as a complaint from detractors of the device. Some governments, including the U.S. and the U.K., vigorously debated legislation to limit the types of privacy violations that the public feared from the device. The device became hampered by public perception and negative publicity. Eventually, Google announced a partial discontinuation of the device in 2015, with full removal of the device from the market in 2023. 

With all of the concerns of video capture and facial recognition aside, this device was relatively limited by modern standards. The Glass had no depth sensors, a limited camera resolution, and restrictive onboard processing power. In contrast, today's MR devices are far more powerful and are becoming capable of the types of exploitation feared in 2013. However, we have yet to see the types of public outcry that plagued Google Glass against current MR devices. Likely, this is because these devices have not yet achieved the types of consumer-grade ubiquitous wear that Glass has sought. Even so, from examining the Glass and its troubles with negative public perception, we can see clear evidence that public scrutiny is high around devices that can record audio or visual data in everyday situations, potentially surreptitiously, and without explicit consent from the persons recorded. The Google Glass example demonstrates a case where the specific technical vulnerability that caused the privacy concerns of bystanders did not exist, still, the significant barrier to product success was a result of negative public perception.

\section{ADDRESSING BYSTANDER PRIVACY ISSUES}
\label{AddressingIssues}

\subsection{Principles}
Any viable design or system that seeks to improve bystander expectations of privacy in the face of modern MR devices is required to maintain an acceptable immersive experience that benefits the device user. This means that practical solutions not only protect the information of bystanders, but also keep from reducing device rendering speeds, introducing latency, adding unnecessary user input (Graphical User Interfaces (GUIs) or physical tokens/actions), and preventing legitimate operations (such as object detection or facial recognition).

\textbf{Usability.} MR devices provide an immersive and fluid experience but have strict requirements for performance. As an example, device frame rates ensure that the digital world is rendered fast enough to ensure the experience does not appear choppy or cause sickness to the user. Bystander privacy solutions that rely on compute-intensive mechanisms such as onboard machine learning inference can interfere with this experience by adding additional overhead to the device, reducing frame rates, and negatively impacting user experience. An optimal solution should have a small enough computational footprint to be completely seamless for the user. Additionally, the solution should not present overly complicated decisions in the form of GUIs, pop-ups, or require users/bystanders to possess or use complicated artifacts such as QR codes, or other physical items. Such interventions can reduce the ``flow'' that is the optimal experience for MR device users~\cite{AA9} or simply create usability issues that negatively affect the viability of such a solution. 

\textbf{Bystander Protection.} It should be evident that a successful bystander privacy solution should seek to protect bystanders in as many ways as possible. However, no solution currently available can claim to be completely effective in all scenarios. Existing solutions seek to identify and classify the persons in the device's field of view. Some solutions force a classification between bystanders and non-bystanders (e.g., subjects) with the assumption that there must always be a subject~\ref{ref:AutoDetectingBystanders}. To improve on this, a solution may be required to understand nuanced situations where a subject may not always exist, or potentially more than one subject may be present. 

Additionally, bystander privacy solutions must not lose sight of their goals when collecting information about bystanders themselves to be used to classify them. To make this point clear, we use OWASP's definition of a privacy violation, making the offloading of sensitive information one of the two requirements for a true privacy violation~\cite{OWASPPrivacy}. Suppose information is collected by the device to identify bystanders but is not removed from the device (including information gathered from the raw data, like facial features or names). In that case, we do not consider this a true violation.

\textbf{Availability of Legitimate Bystander Information.}
No bystander privacy solution can simply strip away the information of all persons in the device's capture radius. Legitimate applications, such as the memory care example in Fig.~\ref{fig:MedicalExample}, can require facial detection or recognition techniques. Removing all potentially identifying information would reduce the experience and usefulness of the device. Any solution needs a mechanism to decide what persons need to be removed and which do not. Later in this section, we create a dichotomy of solutions based on their mechanisms, with Explicit Solutions requiring user input and Implicit Solutions using the available context in the interaction to separate the subject from bystanders.

\textbf{Consent.} 
  From both the case study on Google's Glass, legislation, and proposed policies, two facets of a successful bystander privacy
  system have emerged - bystander awareness of the device's state and bystander consent to be recorded~\ref{ref:KatzvUS}~\ref{ref:PrivacyQuesationsExpansionAR}~\ref{ref:understandingbystandersvaryingneeds}. 

These mechanisms, whether they involve verbal communication
  from the bystander, a physical or digital token (e.g., a cell phone
  with a Near Field Communication (NFC) channel to express recording
  preference and/or consent to nearby devices), or registration with a
  database, have been proven to increase bystander confidence in the
  safety of their personal information in the face of MR devices.

Additionally, consent can be given in different ways.  For instance, if a business decides to limit the use of MR devices in its stores in order to protect bystander privacy, as was the case with Google's Glass, this is a passive consent mechanism. However, we remain optimistic that future solutions would address bystander privacy concerns and limit this blanket-type of passive consent.

\begin{table*}
\centering
\caption{Existing Bystander Privacy Works (in chronological order)} \label{tab:Q2}
\begin{tabular}{|N|p{14cm}|}
\hline
\multicolumn{1}{|c|}{\textbf{Number}} & \textbf{Reference}  \\ \hline
\label{ref:KatzvUS} & U.S. Supreme Court, ``Katz v. United States'', \textit{389 U.S. 347, 350-51}\\
\hline
\label{ref:PrivacyAsContextualIntegrity} & Helen Nissenbaum, ``Privacy as Contextual Integrity'', \textit{Washington Law Review, 79 Wash. L. Rev. 119 (2004)}\\
\hline
\label{ref:ScannerDarkly} & Jana et. al, ``A Scanner Darkly: Protecting User Privacy from Perceptual Applications'', IEEE S\&P '13\\
\hline
\label{ref:FineGrainedPermissionsRecognizers} & Jana et. al., ``Enabling fine-grained permissions for augmented reality applications with recognizers'', USENIX SEC'13\\
\hline
\label{ref:WDAC} & Roesner et. al, ``World-Driven Access Control for Continuous Sensing'', ACM CCS '14\\
\hline
\label{ref:MarkIt} & Raval et. al., ``MarkIt: privacy markers for protecting visual secrets'', ACM UbiComp\\
\hline
\label{ref:InSitu} & Denning et. al., ``In situ with bystanders of augmented reality glasses: perspectives on recording and privacy-mediating technologies'', ACM CHI '14\\
\hline
\label{ref:Cardea} & Shu et. al., ``Cardea: context-aware visual privacy protection for photo taking and sharing'', ACM MMSys '18\\
\hline
\label{ref:PrivacEye} & Steil et. al.,  ``PrivacEye: privacy-preserving head-mounted eye tracking using egocentric scene image and eye movement features'', ACM ETRA '19\\
\hline
\label{ref:AutoDetectingBystanders} & Hasan et. al., ``Automatically Detecting Bystanders in Photos to Reduce Privacy Risks'', IEEE S\&P '20\\
\hline
\label{ref:UserCetnricDesigns} & Ahmad et. al., ``Tangible Privacy: Towards User-Centric Sensor Designs for Bystander Privacy'', ACM HCI '20\\

\hline
\label{ref:PrivacyQuesationsExpansionAR} & Ellysse Dick. 2020. ``How to Address Privacy Questions Raised by the Expansion of Augmented Reality in Public Spaces'', Information Technology and Innovation Foundation\\
\hline
\label{ref:BalancingUserPrivacy} & Ellysse Dick. 2021. ``Balancing User Privacy and Innovation in Augmented and Virtual Reality'', Information Technology and Innovation Foundation\\
\hline
\label{ref:adjacentactorprivacy} & Pierce et. al., ``Addressing Adjacent Actor Privacy: Designing for Bystanders, Co-Users, and Surveilled Subjects of Smart Home Cameras'', ACM DIS '22 \\
\hline
\label{ref:hiddeninplainsight} & Lehman et. al.,  ``Hidden in Plain Sight: Exploring Privacy Risks of Mobile Augmented Reality Applications'', ACM TPS November 2022\\
\hline
\label{ref:understandingbystandersvaryingneeds} & O'Hagan et. al.,``Privacy-Enhancing Technology and Everyday Augmented Reality: Understanding Bystanders' Varying Needs for Awareness and Consent'', IMWUT December 2022\\
\hline
\label{ref:BANS} & Mansour et. al., ``BANS: Evaluation of Bystander Awareness Notification Systems for Productivity in VR'', NDSS Symposium on Usable Security and Privacy '23\\
\hline
\label{ref:BystandAR} & Corbett et. al.,  ``BystandAR: Protecting Bystander Visual Data in Augmented Reality Systems'', ACM MobiSys '23\\
\hline
\end{tabular}

\end{table*}

\subsection{Explicit Solutions}
\label{ExplicitSolutions}
We define explicit solutions to the BPP as systems that require the
device user or the bystander 
to interact with the system to either opt-in or opt-out from the recording by the device's sensors. This can be done through a
published privacy policy, hand gestures, physical tokens, etc.~\ref{ref:BystandAR}. Privacy-policy-based solutions, such as Cardea~\ref{ref:Cardea}, require the bystander to establish a preference profile or upload images to pre-train a
classifier on an edge or cloud server. Such a node then processes this
data to decide if certain portions of the image need to be sanitized
of bystander data by recognizing gestures, faces, sensitive locations,
etc. As another example, Darkly~\ref{ref:ScannerDarkly} requires the device user to manage privacy preference through a GUI but allows bystanders input in the form of hand gestures during capture. Other solutions require specialized physical equipment, actions, or tokens to be worn by either the user or the bystander or require the user to make fine-grained decisions about application permissions without knowing the future context of collection~\ref{ref:FineGrainedPermissionsRecognizers}~\ref{ref:WDAC}~\ref{ref:MarkIt}~\ref{ref:PrivacEye}.

These systems generally provide the bystander with a tangible mechanism of providing their consent to be recorded (e.g., hand gestures), which has been shown to be preferable over implicit systems (i.e., systems that do not allow for bystander input)~\ref{ref:InSitu}~\ref{ref:understandingbystandersvaryingneeds}. However, these systems impose a burden on the bystander to intervene and protect their own privacy, especially with systems that require explicit bystander input in the form of physical tokens or a registration in a system. If a bystander chooses not to do so, they may not be able to expect that the system is working to protect their information from exploitation and misuse. 

\subsection{Implicit Solutions}
\label{ImplicitSolutions}
Implicit BPP systems protect the bystander without explicit actions, by inferring context from the bystander, user, or environment~\ref{ref:BystandAR}. In general, these systems use a machine learning model to infer the presence (or lack) of some indicator to decide if a person is either a subject or a bystander. Some systems, such as BystandAR, use information about the user's eye gaze and voice to determine the subject of an interaction and remove the visual information of the remaining persons in view~\ref{ref:BystandAR}. Others use the position of the person relative to the center of a captured image or the direction of the person's eye gaze~\ref{ref:AutoDetectingBystanders}. 

These systems have the advantage of not requiring explicit input from the bystander to expect protection. By inferring information about the context and environment, the solution does not require hand gestures, tokens, or registration to make a decision about who to protect and when. However, if the context is not as expected, these solutions can falter. For example, if the bystander happens to be captured near the center of the frame, and also happens to be looking at the camera, then a machine learning model can report a false negative for the presence of a bystander. If the solution relies on context from the user, such as in BystandAR, a malicious user could override the protection in order to capture the visual information of victims. 

\subsection{Gaps in Current Solutions} With the exception of only a few (e.g., BystandAR~\cite{BystandAR}), current solutions struggle to operate in real-time on live sensor data. Most existing solutions are designed to protect bystanders after the moment of capture, by offloading and inferring the presence of a bystander on an edge node or similar. This forces such solutions to be used exclusively in roles that can tolerate the delay, such as static image captures for use on social media. When considering MR devices, we recommend that this information be protected in real-time. If not, a legitimate application could not expect access to live data for legitimate reasons (Fig.~\ref{fig:MedicalExample}) without delay. Additionally, transferring unprotected data off the device has been shown to create vulnerabilities during data transmission. This presents a challenge to convincing device users and application designers to integrate bystander privacy solutions in their workflow. 


Additionally, even a perfect technical solution that protects a bystander completely in all cases cannot be deemed successful if bystanders do not perceive the system as safe. As shown by Google's Glass, the perception of the safety of the system can outweigh the actual threat when it comes to marketability and public backlash. While studies such as those done in O'Hagan et. al.~\ref{ref:understandingbystandersvaryingneeds} have illustrated initial privacy directions and issues with public perception, the next step should involve deploying and testing actual bystander protection systems on commodity MR devices and studying the effect they have. 
Much more work is needed to actually gauge the bystander reaction to a real-time, fully implemented system that can be used on MR devices. 

An ideal solution must address the requirements of usability, protection, availability, and consent. The solution must run near-seamlessly on modern MR devices, provide an acceptable amount of protection for bystanders, provide the required information for third-party applications, and give bystanders a mechanism to provide consent to be recorded. Since, to our knowledge, such a system does not yet exist, there remains much work to be done in this regard. 


\section{FUTURE DIRECTIONS IN BYSTANDER PRIVACY}
\label{FutureDirections}
In addition to addressing the lack of a technical solution to address bystander privacy concerns, research must also address the more nuanced areas of public education and perceptions. As shown in previous work, there is simultaneously a lack
of understanding of the threats posed to bystanders, and unnecessary apprehension towards devices that do not pose a threat~\ref{ref:understandingbystandersvaryingneeds}. In order to make future MR devices viable and publicly acceptable, we must address these interrelated problems holistically. 

\subsection{Education}
At the root of the perception problem facing MR devices, education provides an understanding of the true threat of such devices. A lack of either user or bystander education has been shown to be the cause of both the demise of past devices and a lack of understanding of the actual threats posed by modern ones. For instance, the Glass suffered from a misunderstanding of the device's capability to record and exploit data, while modern devices that pose far more risk have been shown to be overlooked~\ref{ref:understandingbystandersvaryingneeds}. From this, it stands to reason that a more thorough understanding of the capabilities of modern MR devices would ameliorate both extremes. Understanding that modern MR devices are capable of facial recognition, weight estimation, and voice recording (among other exploitation avenues) would create the wariness that we believe would be commensurate with the threat. This would also direct the public's focus to risks that actually exist, direct future technical and policy solutions towards more refined language, and prevent overly restrictive limits. 

The mechanism to convey this understanding, however, is not as clear. Clearly, existing methods have failed to impress a full understanding of these threats on the public outside of research circles. There is much room for future research that investigates how best to bridge this understanding gap. Future work should explore what data should be conveyed, at which times, and on what mediums. This information, when properly conveyed, could simultaneously reduce undue scrutiny on unlikely or infeasible threats, while increasing scrutiny on viable ones. 

\subsection{Bystander Perceptions}
Even when provided with the correct information about the viability of a threat, we must better understand the psychological impact of MR devices in public areas. Existing work on this topic focuses on protection techniques and consent mechanisms but does little to evaluate an implemented, real-time system that seeks to provide the proper (and only the proper) information to third-party applications. Currently published surveys on bystander perceptions and apprehensions shed light on this by informing us of bystander tolerances in certain contexts, including when the device is used by friends, strangers, and coworkers. A wider familiarization with MR devices may be necessary to fully realize any solution~\ref{ref:InSitu}. Other surveys explore new ``body-based gestures'' like those available as input mechanisms on some modern MR devices. These studies find that the new, MR-friendly methods are less acceptable than others to the general public~\cite{GestureAcceptability}. Using implemented and tested solutions, such as BystandAR~\ref{ref:BystandAR}, future work can explore the actual impact of technical solutions on bystander perceptions of privacy.

\subsection{Experience}
Certain technologies have always created distrust at the onset of their proliferation in the public. Even the cellphone, now ubiquitous, was viewed warily by a public concerned with its ability to record personal data as recently as the year 2000~\cite{GoingWireless}. However, over time these technologies have become more widely used and more widely accepted. It also follows that as MR technology matures and potentially becomes more ubiquitous, the context of their acceptable use changes as well~\ref{ref:PrivacyAsContextualIntegrity}.

\subsection{Nuance in Labeling Subjects and Bystanders}
Existing solutions generally tend to label subjects and bystanders according to a simple dichotomy. However, as discussed in Section~\ref{se:WhatisABystander}, more detailed and nuanced definitions exist. Potential future solutions could sub-classify subjects as ``co-users'' and subjects, giving more protection to those whom the MR user intends to interact with, as opposed to ``co-users'' who are fully aware and can control the application and data collection~\ref{ref:adjacentactorprivacy}. Additionally, future solutions can protect persons who potentially have given permission for data collection in the past but are not relevant to the current interaction context~\ref{ref:adjacentactorprivacy}.

\section{CONCLUSION}
Bystander concerns about privacy in public situations have been shown to be crucial to the wide acceptance of MR devices. Technical solutions to this suffer from a lack of context or viability on modern MR devices but 
make progress towards easing bystander concern with advances such as explicit consent mechanisms or implicit contextual understanding. However, much work is still needed to fully understand the impact of the bystander privacy problem on future devices, and also in designing more efficient and viable solutions. We believe that these potential future solutions can reduce the public apprehension that has plagued devices in the past, and make the devices that support the Metaverse more prolific and less controversial. If so, we move even closer to the ultimate promise of the Metaverse.

\vspace*{-8pt}

\section{ACKNOWLEDGMENTS}

This work was supported under NSF grants 2112778 and 2153397 and by the Commonwealth Cyber Initiative (CCI). 

\def\refname{REFERENCES}


\bibliographystyle{IEEEtran}
\bibliography{BystanderPrivacy}

\end{document}